\documentclass[twocolumn]{jsiamletters}

\usepackage{color}
\usepackage{ulem}
\usepackage{amsmath,amssymb,amsthm}
\theoremstyle{definition}
\newtheorem{lem}{Lemma}[section]

\theoremstyle{definition}
\newtheorem{defn}{Definition}[section]

\newtheorem{exmp}{Example}[section]

\theoremstyle{definition}
\newtheorem{thm}{Theorem}[section]

\theoremstyle{remark}

\group{Applied Chaos}
\affiliation{Graduate school of Informatics, Kyoto University}{Yoshida-honmachi, Kyoto city, Japan}
\authorinfo{Atsushi Iwasaki}{1}{iwasaki.atsushi.47e@st.kyoto-u.ac.jp}
\authorinfo{Ken Umeno}{1}{ken.umeno.8z@kyoto-u.ac.jp}
\title{One-stroke polynomials over a ring of modulo $2^w$}
\abstract{
Permutation polynomials over a ring of modulo $2^w$ are {well adopted to} digital computers and  digital signal processors, and so they are in particular expected to be useful for cryptography and pseudo random number generator{s}.
{Since a longer period of the polynomial  is demanded in general, we derive a necessary and sufficient condition that polynomials are permutating and their periods are the longest over the ring.
We call polynomials which satisfy the condition ``one-stroke polynomials over the ring".}

}
\begin{document}

\maketitle

\section{Introduction}
A polynomial is called a permutation polynomial over a finite ring $R$ if the polynomial is bijection over $R$.
Although $R$ is a finite field in many studies, we deal with a ring of modulo $2^w$ in this paper.
Studies about permutation polynomials over the ring are very important because they are {well adopted to} with digital computers and  digital signal processors.
They can calculate values of permutation polynomials over the ring faster than over a finite field because 2 power residue operation is practically negligible.
Then, they are in particular expected to be useful for cryptography and pseudo random number generator{s}, and some applications {have been} already proposed \cite{RC6, Umeno-Kim-Hasegawa, Iwasaki-Umeno}.

There are two important studies about permutation polynomials over the ring.
One is about periods of the polynomials.
{For cryptography and pseudo random number generators, such periods are expected to be longer.
Then, a} necessary and sufficient condition to maximize the period{s} of the permutation polynomial{s} should be {explored}.
When the period of the permutation polynomial is maximized, there exists only one orbit {passed} by the polynomial over the ring and the orbit {passes all the elements of} the ring.
Since a map which draws such only one orbit is called ``one-stroke map" \cite{Kyouritsu}, we call such permutation polynomials ``one-stroke polynomials" in this paper.
The necessary and sufficient condition {that} specifies one-stroke polynomials with the assumption that the degree of the permutation polynomials are restricted to 1 or 2 is known \cite{Knuth}.
One-stroke polynomials whose degrees are 1 or 2 are used in {a} linear congruential method and {a} quadratic congruential method, which are  pseudo random number generators.
A sufficient condition without {any} assumption {has also been} known \cite{Coveyou}, but {a} necessary and sufficient condition without the assumption {has not been} known as far as the authors know.

The other is more fundamental.
In order to study about permutation polynomials over a ring of modulo $2^w$, we should know which polynomials are permutation polynomials.
The necessary and sufficient condition {that} specifies permutation polynomials {have been} already studied \cite{Rivest}.

Based on the above, we study about the one-stroke polynomials over a ring of modulo $2^w$ whose degrees are {\it arbitrary}.
This paper is constructed as follows.
In section 2, we introduce permutation polynomials over a ring of modulo $2^w$.
In section 3, we derive the necessary and sufficient condition to specifies one-stroke polynomials over the ring.
In section 4, we introduce some properties about one-stroke polynomials over the ring.
Finally, we conclude this paper.

\section{Permutation polynomials over a ring of modulo $2^w$}

In this section, we introduce permutation polynomials over a ring of modulo $2^w$.

\begin{defn}
A finite degree polynomial $f(X)$ with integer coefficients is called a permutation polynomial over a ring of modulo $2^w$ if 
\begin{align*}
\forall w\geq0,\ \{f(\Bar{X})\mod2^w|\Bar{X}\in\mathbb{Z}/2^w\mathbb{Z}\}=\mathbb{Z}/2^w\mathbb{Z}.
\end{align*}
\end{defn}

The necessary and sufficient condition {that} specifies permutation polynomials over the ring is given by the following theorem \cite{Rivest}.
\begin{thm}
\label{thmRivest}
{\bf [Rivest, 2001] }A polynomial $f(X)=\sum_{i=0}^Na_iX^i$, where the coefficients are integers, is a permutation polynomial over a ring of modulo $2^w$ if and only if
\begin{align}
a_1&\equiv1\mod2,\label{RivestA}\\
(a_2+a_4+a_6+\cdots)&\equiv0\mod2,\label{RivestB}\\
(a_3+a_5+a_7+\cdots)&\equiv0\mod2\label{RivestC}.
\end{align} 
\end{thm}
The following lemma is used in order to prove Theorem \ref{thmRivest}.
We also use the lemma in the next section.

\begin{lem}
\label{lem2-1}
Let $f(X)$ is a polynomial with integer coefficients.
Then, $f(X)$ is a permutation polynomial over a ring of modulo $2^w$ if and only if
\begin{align*}
\forall w\geq1,\ f(X+2^{w-1})\equiv f(X)+2^{w-1}\mod2^w.
\end{align*}
\end{lem}

The following lemma is also used in the next section.

\begin{lem}
\label{lem2-2}
Let $f(X)$ is a permutation polynomial over a ring of modulo $2^w$.
Then, $f^j(X)$ is also a permutation polynomial over the ring for arbitrary integer $j$, where $f^j(X):=f\circ f^{j-1}(X)$ and $f^1(X):=f(X)$. 
\end{lem}

\section{One-stroke polynomial}
In this section, we derive a necessary and sufficient condition {that} coefficients of one-stroke polynomials over a ring of modulo $2^w$ satisfy.
First, we exactly define one-stroke polynomials over a ring of modulo $2^w$.

\begin{figure}[h]

\begin{center}
\begin{picture}(180,68)
\put(0,20){3}
\put(2,30){\vector(0,1){15}}
\put(0,50){0}
\put(10,54){\vector(1,0){15}}
\put(30,50){1}
\put(32,47){\vector(0,-1){15}}
\put(30,20){2}
\put(25,23){\vector(-1,0){15}}
\put(13,5){(a)}

\put(90,20){3}
\put(92,30){\vector(0,1){15}}
\put(90,50){0}
\put(100,54){\vector(1,0){15}}
\put(120,50){1}
\put(122,47){\vector(0,-1){15}}
\put(120,20){2}
\put(115,23){\vector(-1,0){15}}

\put(150,50){4}
\put(160,54){\vector(1,0){15}}
\put(180,50){5}
\put(182,47){\vector(0,-1){15}}
\put(180,20){6}
\put(175,23){\vector(-1,0){15}}
\put(150,20){7}
\put(152,30){\vector(0,1){15}}
\put(133,5){(b)}
\end{picture}
\end{center}
\caption{Orbits passed by $G(X)$. (a) Orbit on $\mathbb{Z}/2^2\mathbb{Z}$. 
(b) Orbit on $\mathbb{Z}/2^3\mathbb{Z}$.}
\label{orbit2}
\end{figure}
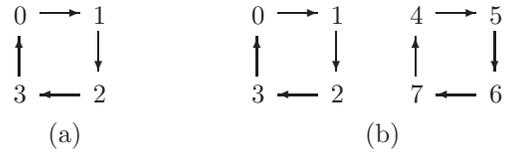

\begin{defn}
\label{def1}
Let $f(X)$ is a permutation polynomial over a ring of modulo $2^w$.
If $f(X)$ satisfy
\begin{align*}
\forall w\geq1,\ {\forall \Bar{X},}\ \{f^i(\Bar{X})\mod2^w|i\in\mathbb{Z}/2^w\mathbb{Z}\}=\mathbb{Z}/2^w\mathbb{Z},
\end{align*}
$f(X)$ is called a one-stroke polynomial over a ring of modulo $2^w$.
\end{defn}
{
\begin{exmp}
We consider polynomials $F(X)=4X^3+X+1$ and $G(X)=6X^3+2X^2+X+1$.
Both of them are permutation polynomials over a ring of modulo $2^w$.
Fig. \ref{orbit1} and \ref{orbit2} show orbits on a ring of modulo $2^w$ passed by $F(X)$  and $G(X)$, respectively.
In Fig. \ref{orbit1}, each orbit passes all elements of the ring where the orbit is passed on.
It means that $F(X)$ is a one-stroke polynomial over a ring of modulo $2^w$.
On the other hand, $G(X)$ is not a one-stroke polynomial over a ring of modulo $2^w$ because there is not an orbit which passes all elements of $\mathbb{Z}/2^3\mathbb{Z}$.  
\end{exmp}
}
\begin{figure}[h]

\begin{center}
\begin{picture}(180,158)
\put(0,20){2}
\put(2,30){\vector(0,1){15}}
\put(0,50){3}
\put(2,60){\vector(0,1){15}}
\put(0,80){0}
\put(10,84){\vector(1,0){15}}
\put(30,80){1}
\put(40,84){\vector(1,0){15}}
\put(60,80){6}
\put(70,84){\vector(1,0){15}}
\put(90,80){7}
\put(100,84){\vector(1,0){15}}
\put(120,80){4}
\put(130,84){\vector(1,0){15}}
\put(150,80){5}
\put(160,84){\vector(1,0){15}}
\put(177,80){10}
\put(182,77){\vector(0,-1){15}}
\put(177,50){11}
\put(182,47){\vector(0,-1){15}}
\put(180,20){8}
\put(175,23){\vector(-1,0){15}}
\put(150,20){9}
\put(145,23){\vector(-1,0){15}}
\put(117,20){14}
\put(115,23){\vector(-1,0){15}}
\put(87,20){15}
\put(85,23){\vector(-1,0){15}}
\put(57,20){12}
\put(55,23){\vector(-1,0){15}}
\put(27,20){13}
\put(25,23){\vector(-1,0){15}}
\put(87,5){(c)}

\put(0,130){3}
\put(2,140){\vector(0,1){15}}
\put(0,160){0}
\put(10,164){\vector(1,0){15}}
\put(30,160){1}
\put(32,157){\vector(0,-1){15}}
\put(30,130){2}
\put(25,133){\vector(-1,0){15}}
\put(13,115){(a)}

\put(90,130){3}
\put(92,140){\vector(0,1){15}}
\put(90,160){0}
\put(100,164){\vector(1,0){15}}
\put(120,160){1}
\put(130,164){\vector(1,0){15}}
\put(150,160){6}
\put(160,164){\vector(1,0){15}}
\put(180,160){7}
\put(182,157){\vector(0,-1){15}}
\put(180,130){4}
\put(175,133){\vector(-1,0){15}}
\put(150,130){5}
\put(145,133){\vector(-1,0){15}}
\put(120,130){2}
\put(115,133){\vector(-1,0){15}}
\put(133,115){(b)}
\end{picture}
\end{center}
\caption{Orbits passed by $F(X)$. (a) Orbit on $\mathbb{Z}/2^2\mathbb{Z}$. 
(b) Orbit on $\mathbb{Z}/2^3\mathbb{Z}$. 
(c) Orbit on $\mathbb{Z}/2^4\mathbb{Z}$.}
\label{orbit1}
\end{figure}
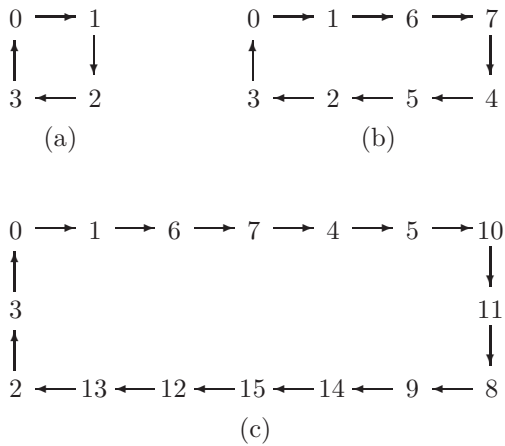

{ Next, we introduce some lemmas.}
By the definition, the following two lemmas are obviously true.

\begin{lem}
\label{lem3-1}
Let $f(X)$ is a permutation polynomial over a ring of modulo $2^w$.
Then, $f(X)$ is a one-stroke polynomial over the ring if and only if
\begin{align*}
f^{i}(0)&\equiv0\mod2^w \ \Leftrightarrow\ i\equiv0\mod2^w.
\end{align*}
\end{lem}

\begin{lem}
\label{lem3-1-2}
Let $f(X)$ is a permutation polynomial over a ring of modulo $2^w$.
Then, $f(X)$ is a one-stroke polynomial over the ring if and only if
\begin{align*}
f^{2^w}(0)&\equiv0\mod2^w,\\
f^{2^{w-1}}(0)&\not\equiv0\mod2^w.
\end{align*}
\end{lem}

\begin{lem}
\label{lem3-2}
Let $f(X)$ is a permutation polynomial over a ring of modulo $2^w$.
Then, $f(X)$ is a one-stroke polynomial over the ring if and only if
\begin{align}
\label{lem3-2A}
\forall w\geq1,\ f^{2^{w-1}}(0)\equiv 2^{w-1}\mod2^w.
\end{align}
\end{lem}

\noindent{\bf Proof}
Assume that $f(X)$ is a one-stroke polynomial over the ring.
By the definition,
\begin{align*}
\forall w\geq1,\ \exists i\leq2^w,\ \text{s.t.}\ f^i(0)\equiv 2^{w-1}\mod2^w.
\end{align*}
Then, by Lemma \ref{lem2-1} and \ref{lem2-2},
\begin{align*}
f^{2i}(0)\equiv f^i(2^{w-1})\mod2^w\equiv0\mod2^w.
\end{align*}
By Lemma \ref{lem3-1},\ $2i=2^w$.
Then, $i=2^{w-1}$.

Conversely, assume that (\ref{lem3-2A}) is true.
Then, by Lemma \ref{lem2-1} and \ref{lem2-2},
\begin{align*}
f^{2^w}(0)\equiv f^{2^{w-1}}(2^{w-1})\mod2^w\equiv 0\mod2^w.
\end{align*}
By Lemma \ref{lem3-1-2}, $f(X)$ is a one-stroke polynomial over the ring.\qed
\vspace{3mm}

\begin{lem}
\label{lem3-3}
Assume that $f(X)$ is a permutation polynomial over a ring of modulo $2^w$ and $f(X)$ satisfy $f^2(0)\equiv2\mod4$ and $f^4(0)\equiv4\mod8$.
Then,
\begin{align*}
\forall w\geq2,\ f^{2^{w-1}}(0)\equiv2^{w-1}\mod2^w.
\end{align*}
\end{lem}

\noindent{\bf Proof}
Assume that $f^2(X)=\sum b_iX^i$ and $f^4(X)=\sum c_iX^i$, where all $b_i$ and $c_i$ are integers.
By the assumption of the lemma, $b_0\equiv2\mod4$ and $c_0\equiv4\mod8$.
Since $f(X)$ is a permutation polynomial over the ring, by Lemma \ref{lem2-2}, $f^2(X)$ is also permutation polynomial over the ring.
Then, by the Theorem \ref{thmRivest}, $b_1\equiv1\mod2$.
Since $f^4(X)=f^2\circ f^2(X)$,
\begin{align*}
c_1
=&b_1^2+2b_2b_1b_0+3b_3b_1b_0^2+4b_4b_1b_0^3+\cdots\\
\equiv& b_1^2\mod4\ \  (\because b_0\equiv2\mod4)\\
\equiv& 1\mod4\ \ (\because b_1\equiv1\mod2).
\end{align*}

Assume that there exists an integer $\Bar{w}\geq3$ such that $f^{2^{\Bar{w}-1}}(0)\equiv2^{\Bar{w}-1}\mod2^{\Bar{w}}$ and the first degree's coefficient of the $f^{2^{\Bar{w}-1}}(X)$ is 1 under modulo 4.
We express $f^{2^{\Bar{w}-1}}(X)$ and $f^{2^{\Bar{w}}}(X)$ as $f^{2^{\Bar{w}-1}}(X)=\sum d_iX^i$ and $f^{2^{\Bar{w}}}(X)=\sum e_iX^i$, where all $d_i$ and $e_i$ are integers.
By the assumption, $d_1\equiv1\mod4$ and $d_0\equiv2^{\Bar{w}-1}\mod2^{\Bar{w}}$.
\begin{align*}
e_1=&d_1^2+2d_2d_1d_0+3d_3d_1d_0^2+4d_4d_1d_0^3+\cdots\\
\equiv& d_1^2\mod4\ \  (\because d_0\equiv2^{\Bar{w}-1}\mod2^{\Bar{w}})\\
\equiv& 1\mod4\ \ (\because d_1\equiv1\mod2),\\
e_0=&d_0+d_1d_0+d_2d_0^2+d_3d_0^3+\cdots\\
\equiv& d_0+d_0d_1\mod2^{\Bar{w}+1}\ \  (\because d_0\equiv2^{\Bar{w}-1}\mod2^{\Bar{w}})\\
\equiv& 2^{\Bar{w}}\mod2^{\Bar{w}+1}\ \ (\because d_1\equiv1\mod4).
\end{align*}
Then, $f^{2^{\Bar{w}}}(0)\equiv2^{\Bar{w}}\mod2^{\Bar{w}+1}$ and the first degree's coefficient of $f^{2^{\Bar{w}}}(X)$ is 1 under modulo 4.

From the above, the lemma is true.\qed

{ Finally, we introduce a necessary and sufficient condition that specifies one-stroke polynomials over a ring of modulo $2^w$.}
\begin{thm}
\label{one-stroke}
Let $f(X)=\sum_{i=0}^Na_iX^i$ is a polynomial, where all $a_i$ are integers.
Then, $f(X)$ is a one-stroke polynomial over a ring of modulo $2^w$ if and only if
\begin{align*}
a_0\equiv&1\mod2, \\
a_1\equiv&1\mod2,\\
(a_2+a_4+a_6+\cdots)\equiv&0\mod2,\\
(a_3+a_5+a_7+\cdots)\equiv&2a_2\mod4,\\
(a_1+a_2+a_3+\cdots)\equiv&1\mod4.
\end{align*}
\end{thm}

\noindent{\bf Proof}
If $f(X)$ is a one-stroke polynomial over the ring, $f(X)$ is a permutation polynomial over the ring.
Then, by Theorem \ref{thmRivest},  Lemmas \ref{lem3-2} and \ref{lem3-3}, $f(X)$ is a one-stroke polynomial over the ring if and only if (\ref{RivestA}), (\ref{RivestB}), (\ref{RivestC}) and  
\begin{align*}
f(0)\equiv1\mod2,\ 
f^2(0)\equiv2\mod4,\ 
&f^4(0)\equiv4\mod8.
\end{align*}

Since $f(0)=a_0$, 
\begin{align*}
f(0)\equiv1\mod2\Leftrightarrow a_0\equiv1\mod2.
\end{align*}

Since $f^2(0)=a_0+a_1a_0+a_2a_0^2+\cdots+a_Na_0^N$, if $a_0\equiv1\mod2$,  (\ref{RivestA}) and (\ref{RivestC}),
\begin{align*}
f^2(0)
\equiv& a_0(1+a_1+a_3+a_5+\cdots)\\
&\ \ \ \ \ \ \ \ \ +(a_2+a_4+a_6+\cdots)\mod4\\
\equiv&1+a_1+a_2+a_3+\cdots+a_N\mod4.
\end{align*}
Then,
\begin{align*}
f^2(0)\equiv2&\mod4,\ a_0\equiv1\mod2,\ (\ref{RivestA})\text{ and }(\ref{RivestC})\\
\Leftrightarrow (a_1+a_2&+a_3+\cdots)\equiv1\mod4,\\ &\ \ a_0\equiv1\mod2,\ (\ref{RivestA})\text{ and }(\ref{RivestC}).
\end{align*}

We express $f^2(X)$ as $f^2(X)=\sum b_iX^i$, where all $b_i$ are integers.
If $f(X)$ is a permutation polynomial over the ring, $f^2(X)$ is also a permutation polynomial over the ring by Lemma \ref{lem2-2}, and so $b_1\equiv1\mod2$ by Theorem \ref{thmRivest}.
If $b_0\equiv2\mod4$ and $b_1\equiv1\mod2$,
\begin{align*}
f^4(0)=&b_0+b_1b_0+b_2b_0^2+b_3b_0^3+\cdots\\
\equiv&2(1+b_1+2b_2)\mod8.
\end{align*}
If $a_0\equiv1\mod2$, (\ref{RivestA}), (\ref{RivestB}) and (\ref{RivestC}),
\begin{align*}
b_2=&a_2a_1+\sum_{i=2}^Na_i\left\{\frac{i(i-1)}{2}a_1^2a_0^{i-2}+ia_2a_0^{i-1}\right\}\\
\equiv& a_2+\sum_{i=2}^Na_i\left\{\frac{i(i-1)}{2}+ia_2\right\}\mod2\\
\equiv& a_2+\sum_{i=2}^Na_i\left\{\frac{i(i-1)}{2}\right\}\mod2\ \ (\because (\ref{RivestC}))\\
\equiv& a_2+(a_3+a_7+a_{11}+\cdots)\\&\ \ \ \ \ \ \ \ +(a_2+a_6+a_{10}+\cdots)\mod2\\
\equiv& (a_3+a_7+a_{11}+\cdots)+(a_6+a_{10}+a_{14}\cdots)\mod2,\\
b_1=&a_1^2+2a_2a_1a_0+3a_3a_1a_0^2+\cdots+Na_Na_1a_0^N\\
\equiv& 1+a_1(3a_3+5a_5+7a_7+\cdots)\\&\ \ \ \ \ \ \ \ +a_1a_0(2a_2+4a_4+6a_6\cdots)\mod4\\
\equiv& 1+a_1(3a_3+a_5+3a_7+\cdots)\\&\ \ \ \ \ \ \ \ +a_1a_0(2a_2+2a_6+2a_{10}\cdots)\mod4\\
\equiv&1+2a_1(a_3+a_7+a_{11}\cdots)+a_1(a_3+a_5+a_7+\cdots)\\&\ \ \ \ \ \ \ \ +2(a_2+a_6+a_{10}\cdots)\mod4\\
\equiv&1+2a_2+2(a_3+a_7+a_{11}+\cdots)\\& \ \ +(a_3+a_5+a_7+\cdots)\\&\ \ \ \ +2(a_6+a_{10}+a_{14}\cdots)\mod4.
\end{align*}
Then,
\begin{align*}
f^4(0)\equiv4+2\{2a_2+(a_3+a_5+a_7+\cdots)\}\mod8.
\end{align*}
Therefore,
\begin{align*}
&f^4(0)\equiv4\mod8, b_0\equiv2\mod4, (\ref{RivestA}), (\ref{RivestB})\text{ and }(\ref{RivestC})\\
\Leftrightarrow& (a_3+a_5+a_7+\cdots)\equiv2a_2\mod4,\\&\ \ \ \ \ \ \ \ \ \ \ \ \ \ \ \ \ b_0\equiv2\mod4, (\ref{RivestA}), (\ref{RivestB})\text{ and }(\ref{RivestC}).
\end{align*}

From the above, the theorem is true.\qed


\section{Some properties of one-stroke polynomials}

In this section, we introduce some properties of one-stroke polynomials.
We show computability of one-stroke polynomials.
Under the assumption that the degree of one-stroke polynomial $f(X)$ is lower than $w$, we show that following values can be calculate with polynomial order times of $w$.
\renewcommand{\labelenumi}{(\Alph{enumi})}
\begin{enumerate}
\item $\Bar{X}$ satisfying $\Bar{Y}\equiv f(\Bar{X})\mod2^w$ for given $\Bar{Y}$.
\item $j$ satisfying $\Bar{Y}\equiv f^j(\Bar{X})\mod2^w$ for given $\Bar{X}$ and $\Bar{Y}$.
\item $\Bar{Y}$ satisfying $\Bar{Y}\equiv f^j(\Bar{X})\mod2^w$ for given $\Bar{X}$ and $j$.
\end{enumerate}
In the paper \cite{Iwasaki-Umeno-arxiv}, similar problem for permutation polynomials over the ring is discussed.
Here, we use not only properties of permutation polynomials over the ring but also those of one-stroke polynomials over the ring.

\noindent{\bf Method to calculate (A).}
The following algorithm can calculate (A).
\renewcommand{\labelenumi}{(\roman{enumi})}
\begin{enumerate}
\item Set $X^\prime\leftarrow0$ and $m\leftarrow1$.
\item If $\Bar{Y}\not\equiv f(X^\prime)\mod2^m$, $X^\prime\leftarrow2^{m-1}$.
\item If $m=w$, output $X^\prime$ and finish this algorithm.
Else, $m\leftarrow m+1$ and return to (ii).
\end{enumerate}
In the step (ii), if $\Bar{Y}\equiv f(X^\prime)+2^{m-1}\mod2^m$, $\Bar{Y}\equiv f(X^\prime+2^{m-1})\mod2^m$ by Lemma \ref{lem2-1}.
Therefore, this algorithm can calculate (A).

Since the degree of $f(X)$ is lower than $w$, it requires $O(w)$ multiplications and $O(w)$ additions on $\mathbb{Z}/2^w\mathbb{Z}$ to calculate the value of $f(X)\mod2^w$ for given $X$.
Thus, the calculation requires $O(w^3)$ times.
Since the calculation is used $O(w)$ times in the above algorithm, the above algorithm requires $O(w^4)$ times.


\noindent{\bf Method to calculate (B).}
In order to calculate (B), we introduce polynomials $h_{2^i}(X)\ (i=0,1,2,\cdots,w-1)$ described as
\begin{align*}
h_{2^i}(X):=\left(f^{2^i}(X)\mod2^w\right)\mod X^{\lceil\frac{w}{i}\rceil}.
\end{align*}
The polynomials $h_{2^i}(X)$ have the following properties.
If $\Bar{X}\equiv 0\mod2^i$,
\begin{align*}
h_{2^i}(\Bar{X})\equiv f^{2^i}(\Bar{X})\mod2^{w}.
\end{align*}
If $\Bar{X}\equiv 0\mod2^{i+1}$,
\begin{align*}
h_{2^i}(\Bar{X})\equiv 2^i\mod2^{i+1},
\end{align*}
and if $\Bar{X}\equiv 2^i\mod2^{i+1}$,
\begin{align*}
h_{2^i}(\Bar{X})\equiv 0\mod2^{i+1}.
\end{align*}
If we know $h_{2^i}(X)$, we can calculate $h_{2^{i+1}}(X)$ as $h_{2^{i+1}}(X)=h_{2^i}\circ h_{2^i}(X)\mod X^{\lceil\frac{w}{i+1}\rceil}$.
Because the degrees of $h_{2^i}(X)$ and $h_{2^{i+1}}(X)$ are lower than $\lceil\frac{w}{i}\rceil$, the calculation requires $O(\lceil\frac{w}{i}\rceil^3)$ multiplications and $O(\lceil\frac{w}{i}\rceil^3)$ additions.
Then, the calculation requires $O(w^2\lceil\frac{w}{i}\rceil^3)$ times.

By the estimation, it takes $O(w^5)$ times to calculate the list $\{h_{2^0}(X),\ h_{2^1}(X),h_{2^2}(X),\ \cdots,\ h_{2^{w-1}}(X)\}$.

We show a method to calculate (B) by using $h_{2^i}(X)$.
If we find $j^\prime$ and $j^{\prime\prime}$ such that
\begin{align*}
0\equiv f^{j^\prime}(\Bar{Y})\mod2^w\ \text{and}\ 0\equiv f^{j^{\prime\prime}}(\Bar{X})\mod2^w,
\end{align*}
we can calculate as $j\equiv j^{\prime\prime}- j^\prime\mod2^w$. 
We, therefore, assume that $\Bar{Y}$ equals to 0 without loss of generality.
Assume that $j=\sum_{i=0}^{w-1}\epsilon(i)2^i$ where $\epsilon(i)\in\{0,1\}$.
Then, $f^j(\Bar{X})\equiv f^{\epsilon(w-1)2^{w-1}}\circ f^{\epsilon(w-2)2^{w-2}}\circ \cdots \circ f^{\epsilon(0)2^0}(\Bar{X})\mod2^w$.
By Lemma \ref{lem3-1},
if $f^j(X)\equiv 0\mod2^w$,
then $f^{\epsilon(m)2^{m}}\circ f^{\epsilon(m-1)2^{m-1}}\circ \cdots \circ f^{\epsilon(0)2^0}(\Bar{X})\equiv0\mod2^{m+1}$ for arbitrary $m$.
Thus, by the properties of $h_{2^i}(X)$,
\begin{align*}
f^j(\Bar{X})\equiv h_{2^{w-1}}^{\epsilon(w-1)}\circ h_{2^{w-2}}^{\epsilon(w-2)}\circ \cdots h_{2^0}^{\epsilon(0)}\mod2^w. 
\end{align*}
From the above, the following algorithm outputs $j$ satysfing $f^j(\Bar{X})\equiv0\mod2^w$.
\begin{enumerate}
\item Set $i\leftarrow0$, $j\leftarrow0$ and $X^\prime=\Bar{X}$.
\item If $X^\prime\equiv2^i\mod2^{i+1}$, $X^\prime\leftarrow h_{2^i}(X^\prime)\mod2^w$ and $j\leftarrow j+2^i$.
\item If $i=w-1$, output $j$ and finish this algorithm.
Else, $i\leftarrow i+1$ and return to step 2.
\end{enumerate}
It takes $O(w^2\lceil\frac{w}{i}\rceil)$ times to calculate the value of $h_{2^i}(\Bar{X})$ for given $\Bar{X}$.
Then, this algorithm requires $O(w^3\log w)$ times, but calculating (B) requires $O(w^5)$ because we must calculate the list $\{h_{2^0}(X),\ h_{2^1}(X),h_{2^2}(X),\ \cdots,\ h_{2^{w-1}}(X)\}$.

\noindent{\bf Method to calculate (C).}
By using the above algorithm, we can find $j^\prime$ such that $f^{j^\prime}(\Bar{X})\equiv0\mod2^w$, and so 
it is enough to show an algorithm to calculate $f^k(0)\mod2^w$ for given $k$.
Assume that $k=\sum_{i=0}^{w-1}\epsilon(i)2^i$ where $\epsilon(i)\in\{0,1\}$.
Then, $f^k(0)\equiv f^{\epsilon(0)2^0}\circ f^{\epsilon(1)2^1}\circ \cdots \circ f^{\epsilon(w-1)2^{w-1}}(0)\mod2^w$.
By Lemma \ref{lem3-1},
$f^{\epsilon(m)2^{m}}\circ f^{\epsilon(m+1)2^{m+1}}\circ \cdots \circ f^{\epsilon({w-1})2^{w-1}}(0)\equiv0\mod2^{m+1}$ for arbitrary $m$.
Thus, by the properties of $h_{2^i}(X)$,
\begin{align*}
f^k(0)\equiv h_{2^0}^{\epsilon(0)}\circ h_{2^{1}}^{\epsilon(1)}\circ \cdots h_{2^{w-1}}^{\epsilon(w-1)}\mod2^w. 
\end{align*}
Then, the following algorithm outputs $f^k(0)\mod2^w$.
\begin{enumerate}
\item Set $i\leftarrow w-1$, $X^\prime=0$.
\item If $(i+1)$-th least significant bit of $k$ is $1$, $X^\prime\leftarrow h_{2^i}(X^\prime)\mod2^w$.
\item If $i=0$, output $X^\prime$ and finish this algorithm.
Else, $i\leftarrow i-1$ and return to step 2.
\end{enumerate}
This algorithm also requires $O(w^3\log w)$ times, but calculating (C) requires $O(w^5)$ by the same reason why the method to calculate (B) requires $O(w^5)$ times.

\section{Conclusion}
We derived the necessary and sufficient condition to specify one-stroke polynomials over a ring of modulo $2^w$.
The condition enables us to construct many long sequences with maximum periods such that the distribution of points of the sequences are uniform over the ring.
In addition, one-stroke polynomials have some interesting properties.
One-stroke polynomials will be applied for many fields including cryptography and pseudo random number generator{s}.

\references

\end{document}